# A Closer Look at Edema Area Segmentation in SD-OCT Images Using Adversarial Framework

Yuhui Tao, Yizhe Zhang, Qiang Chen

School of Computer Science and Engineering, Nanjing University of Science and Technology, Nanjing, China

**Abstract.** The development of artificial intelligence models for macular edema (ME) analysis always relies on expert-annotated pixel-level image datasets which are expensive to collect prospectively. While anomaly-detection-based weakly-supervised methods have shown promise in edema area (EA) segmentation task, their performance still lags behind fully-supervised approaches. In this paper, we leverage the strong correlation between EA and retinal layers in spectral-domain optical coherence tomography (SD-OCT) images, along with the update characteristics of weakly-supervised learning, to enhance an off-the-shelf adversarial framework for EA segmentation with a novel layer-structure-guided post-processing step and a test-time-adaptation (TTA) strategy. By incorporating additional retinal layer information, our framework reframes the dense EA prediction task as one of confirming intersection points between the EA contour and retinal layers, resulting in predictions that better align with the shape prior of EA. Besides, the TTA framework further helps address discrepancies in the manifestations and presentations of EA between training and test sets. Extensive experiments on two publicly available datasets demonstrate that these two proposed ingredients can improve the accuracy and robustness of EA segmentation, bridging the gap between weakly-supervised and fully-supervised models.

## 1 Introduction

Spectral-domain optical coherence tomography (SD-OCT) imaging technology has emerged as a valuable tool for diagnosing retinal disorders and analyzing macular edema (ME). By capturing a series of cross-sectional B-scans, SD-OCT allows for fast and noninvasive acquisition of high-resolution 3D images of retina. ME, characterized by the accumulation of fluid and the subsequent swelling of the central retina, is closely associated with retinal disorders, causing moderate to severe vision loss [1]. Therefore, analyzing ME holds significant diagnostic value in identifying related diseases.

With the rise of deep learning (DL), many DL-based methods have been proposed to automatically segment macular fluids. To reduce false negatives in fluid segmentation methods, literatures designed boundary-aware loss functions for model training [2]. Other works proposed using attention mechanisms and multi-scale information to dynamically fit the varying sizes of fluids [3, 4, 5]. In fact, accurately delineating the contour of a certain fluid (e.g., intraretinal fluid (IRF), subretinal fluid (SRF), pigment epithelial detachment (PED), shown in Figure 1) can be difficult in some cases. To avoid the potential disagreement in inter-center annotation [1], as magenta area shown in Figure 1, Fang et al. [6] defined a novel lesion region with a relatively clear boundary, namely edema area (EA), whose upper and lower boundaries are defined as the inner limiting membrane (ILM) and Bruch's membrane (BM) layers, respectively.

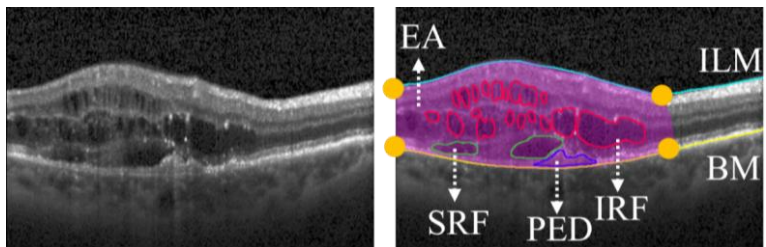

**Fig. 1.** Illustration of EA and three common macular fluids, i.e., SRF, PED, and IRF.

While the above supervised models can segment macular fluid well [2, 3, 4, 5, 6], these methods require labor-intensive annotation which are expensive to collect. To alleviate this limitation, several unsupervised [7, 8, 9, 10] and weakly supervised [11, 12] anomaly detection methods for SD-OCT images have been proposed. These methods mainly adopt the self-reconstruction paradigm where underlying pseudo-healthy counterparts can be reconstructed from pathological input images using a well-trained generator, and then lesions can be detected by calculating the difference between the input and reconstructed images. Among these methods, LAGAN [12] is one of the most competitive models. Since there exists a certain correlation between the retinal layer structure and ME, as ME can cause varying degrees of elevation in the retinal layers. In this paper, we try to further refine the automatic segmentation results of LAGAN with the introduction of additional retinal layer information.

Recently, Test-Time Adaptation (TTA) [13-18] has emerged as a novel research topic in machine learning domain. Its core idea is to design an objective function that does not require manual labels during testing to further optimize the performance of model on test samples. Since ME exhibits diverse manifestations and pixel-level annotations are not necessary in weakly-supervised learning paradigm which is naturally suitable for TTA framework, we investigate in adjusting the model parameters during the testing process to improve the accuracy of EA segmentation.

Overall, the contributions of this work are:

(1) We extend LAGAN with dedicatedly designed layer-structure-guided post-processing, which transforms the dense EA segmentation task into a task of confirming intersection points between EA and retinal layers, producing predictions better conform to the shape prior of EA.

(2) We attempt to adopt TTA strategy in weakly-supervised learning paradigm to adjust the trained model to fit the distribution of EA in test samples.

(3) We conduct extensive experiments on two publicly available datasets and discuss the impact of different strategies in depth. The experiments demonstrate that our framework effectively improves the accuracy of EA segmentation compared with baseline.

## 2 Methods

The original post-processing of LAGAN involves several procedures. Based on the prior knowledge of the correlation between retinal layer structure and ME, we propose a novel layer-structure-guided post-processing to refine the results output by LAGAN. Besides, TTA strategy is applied to further enhance the model's ability to segment EA. The proposed test-time performance improvement method is shown in Algorithm 1.

### 2.1 Layer-structure-guided post-processing

Firstly, an automatic layer segmentation method [19] is employed to obtain the segmentation results of ILM and BM layers. Then, with the guidance of ILM and BM (i.e., the top and bottom boundaries of EA are already known), it is only necessary to confirm the four intersection points (i.e., top-left, bottom-left, top-right, and bottom-right, as orange dots shown in Figure 1, which can define the left and right boundaries of EA) between the predicted contour and retinal layers to obtain the predictions of EA. However, as illustrated in Figure 2, several issues may arise in practice:

(1) Mis-segmentation of BM layer. Some retinal lesions can cause elevation of the retinal pigment epithelium (RPE) layer (first column in Figure 2), and the automatic segmentation model may mistakenly segment the elevated RPE layer as BM layer.

(2) Missing intersection points. Some cases of mild ME do not cause elevation of the ILM layer (second column in Figure 2), leading to the model's predicted upper boundary being below ILM layer. Similarly, some cases of diffuse edema do not significantly alter the retinal tissue (third column in Figure 2), causing the predicted lower boundary deviated from BM layer. This results in the problem of missing intersection points (orange cross in Figure 2) on the same side in the horizontal direction.

(3) Violation of the prior shape of the EA. The left and right boundaries of the EA generally follow a vertical direction, but the segmentation results obtained from the convex hull operation based on the residual map in LAGAN exhibit a significant horizontal discrepancy in the same-side intersection points ($\Delta x$, first column in Figure 2), which contradicts the prior knowledge of the shape of the EA.

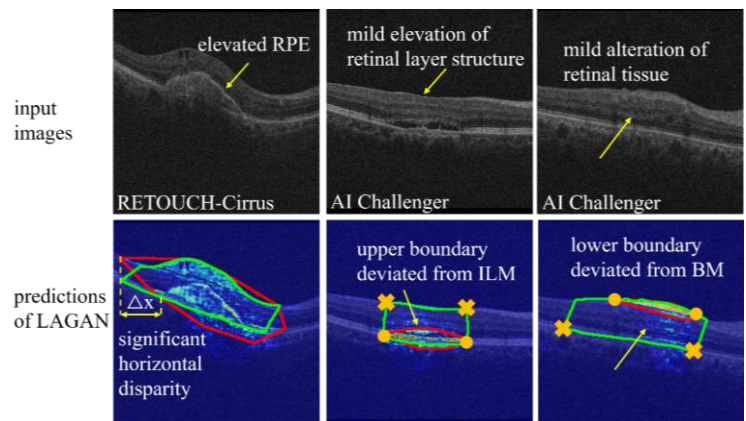

**Fig. 2.** Examples of issues encountered during post-processing based on layer structural constraints. The green and red contours denote the ground truth of EA and the prediction.

To overcome the potential problems caused by issue (1), considering that the topology of the BM layer is generally convex, we propose to use a convex hull fitting strategy to refine the erroneous BM layer segmentation. Additionally, to address the issue (2), the horizontal ordinate of the corresponding intersection point on the same side is selected as horizontal ordinate of the layer's missing intersection point. Lastly, to deal with the issue (3), we propose determining the left and right boundary points of the EA first, and then constructing vertical lines through these boundary points as the left and right boundaries of the prediction.

The specific methods for determining the left and right boundary points can be divided into the following three strategies: Strategy 1/2/3 selects the intersection point that is to the left/right/average of the top-left and bottom-left intersection points as the left boundary point, and the intersection point that is to the right/left/average of the top-right and bottom-right intersection points as the right boundary point, respectively. In this paper, strategy 1 is chosen as the final approach, and the merits and drawbacks of these three strategies will be discussed in section 3.2.

---

**Algorithm 1: Layer-structure-guided post-processing and TTA strategy**

**Input:** trained generator G, trained layer segmentation model S, test image sequence $\{x_t\}_{t=1}^{T}$

**Output:** EA predictions $\{\bar{y}_t\}_{t=1}^{T}$

1:   **for** t = 1, 2, …, T **do**
2:     # **Step 1**. layer-structure-guided post-processing
3:     $ILM, BM \leftarrow S(x_t)$, $BM \leftarrow$ ConvexHull($BM$)     # address issue (1)
4:     $OriSeg \leftarrow$ baseline($G(x_t)$-$x_t$)     # obtain prediction from baseline
5:     traverse to determine the intersection points between $ILM\&BM$ and $OriSeg$
6:     **if** any intersection point does not exist
7:       assign the horizontal ordinate of the other intersection point on the same side to it     # address issue (2)
8:     **end if**
9:     $w_{left} \leftarrow \min(point_{topleft}.x, point_{bottomleft}.x)$     # address issues (3)
10:    $w_{right} \leftarrow \max(point_{topright}.x, point_{bottomright}.x)$     # address issue (3)
11:    $\bar{y}_t \leftarrow$ Confined($ILM, BM, w_{left}, w_{right}$)     # predict EA based on top, bottom, left, and right boundaries
12:    # **Step 2**. TTA strategy
13:    adopting the same loss function proposed in baseline and using $\bar{y}_t$ as pseudo label to update G
14:   **end for**
15:   **return** $\{\bar{y}_t\}_{t=1}^{T}$

---

## 2.2 TTA strategy

LAGAN is trained with the combination of supervised learning (being lesion-aware with the aid of pseudo labels) and adversarial training (for image generation) [20] without utilizing manually annotations. Therefore, the same model update strategy can be adopted to implement TTA. The pseudo-labels used for TTA framework are derived from the model based on LAGAN with the layer constraint strategy (strategy 1 specifically, as explained in section 3.2). For the sake of training stability, only the generator is updated while the discriminator is fixed. Since the learning rate controls the responsiveness of model to the designed loss function during the optimization process and adjusting the learning rate is straightforward. Therefore, we focus on discussing effects of different learning rates on TTA.

In this paper, we consider a more challenging online update scenario: during the testing phase, the model is updated based on pseudo labels, and each test sample is only sent to the model once with a single iteration. Therefore, to prevent the model from catastrophic forgetting, the learning rate needs to be set to a small value. We select from four common learning rates {5e-4, 1e-4, 5e-5, 1e-5} used for model fine-tuning.

## 3 Experiments

### 3.1 Settings

Because this work is an extended version of LAGAN, we use the same experiment setting as illustrated in LAGAN. Readers can refer to reference [12] for more details.

Two publicly available datasets, i.e., AI Challenger[1] and RETOUCH [1], are used in the experiments. We focus on the algorithm aimed at improving the performance of model during the testing process, so we only use the test sets in these two datasets. The AI Challenger test dataset comprises 10 OCT volumes acquired with Cirrus (Zeiss), and each volume contains 128 B-scans. The RETOUCH includes three test datasets from different OCT devices, with 14 volumes acquired with Cirrus, Spectralis (Heidelberg), and T-1000/T-2000 (Topcon), respectively. Cube from Spectralis and Topcon has 49 and 128 B-scans.

To evaluate segmentation performance, we adopt dice coefficient (DSC), intersection over union (IoU), false negative rate (FNR) and false positive rate (FPR).

### 3.2 Results and analysis

The proposed method is compared with LAGAN [12], MPB-CNN++ [6], and two other U-Net [21] variants, i.e., AttU-Net [22] and U-Net++ [23]. Besides, the mainstream TTA methods are not suitable for the adversarial learning framework for EA segmentation in this paper, because they either require additional network architecture design to implement self-supervised tasks [13, 16] or utilize entropy-minimization-based methods in supervised models [14, 15, 17]. Therefore, the proposed TTA framework is not compared with other TTA methods. We term LAGAN and LAGAN + strategy 1 as baseline 1 and baseline 2 for clarity.

---

[1] https://www.challenger.ai

**Table 1.** EA segmentation results (mean ± standard deviation) of different models. The best results are labeled in **bold** and the second best are indicated using underline.

| Dataset | Model | DSC(%)↑ | IoU(%)↑ | FNR(%)↓ | FPR(%)↓ |
|---|---|---|---|---|---|
| AI Challenger | LAGAN | 84.45±11.77 | 74.60±14.99 | 13.94±14.07 | 11.46±12.84 |
| | +strategy 1 | **89.21±8.84** | **81.50±12.32** | **8.90±9.82** | 9.61±11.89 |
| | +strategy 2 | 85.13±11.55 | 75.66±15.57 | 19.09±16.26 | **5.25±8.53** |
| | +strategy 3 | 87.88±9.07 | 79.41±12.81 | 13.36±12.34 | 7.22±10.26 |
| | MPB-CNN++ | 90.57±8.80 | 83.80±12.83 | 8.66±7.63 | 7.54±12.09 |
| | AttU-Net | 91.95±9.86 | 86.22±12.58 | 9.64±12.34 | 4.14±6.09 |
| | U-Net++ | 92.25±7.70 | 86.39±10.97 | 7.50±8.14 | 6.11±9.72 |
| RETOUCH Cirrus | LAGAN | 74.84±16.18 | 62.02±17.48 | 18.15±15.41 | 19.83±19.94 |
| | +strategy 1 | **82.89±14.95** | **72.98±17.63** | **13.12±11.34** | 13.90±18.29 |
| | +strategy 2 | 77.99±15.91 | 66.24±18.19 | 25.11±17.12 | **8.65±16.11** |
| | +strategy 3 | 81.40±14.63 | 70.67±17.00 | 18.27±13.48 | 11.05±17.15 |
| | MPB-CNN++ | 80.40±18.73 | 70.44±20.99 | 22.77±19.23 | 6.79±12.04 |
| | AttU-Net | 74.65±28.99 | 66.09±28.68 | 31.51±29.26 | 2.40±4.55 |
| | U-Net++ | 81.01±21.46 | 72.19±23.26 | 23.46±23.39 | 4.34±7.77 |
| RETOUCH Spectralis | LAGAN | 71.64±24.75 | 60.75±25.98 | 18.45±18.16 | 20.79±25.98 |
| | +strategy 1 | **77.14±21.80** | **67.06±24.41** | **12.31±14.15** | 20.63±25.00 |
| | +strategy 2 | 72.95±24.47 | 62.36±26.15 | 23.98±19.34 | **13.65±23.15** |
| | +strategy 3 | 76.30±22.68 | 66.19±25.02 | 16.87±16.03 | 16.94±24.18 |
| | MPB-CNN++ | 78.04±23.22 | 68.69±25.23 | 19.77±20.93 | 11.54±16.26 |
| | AttU-Net | 83.40±21.21 | 75.78±23.86 | 15.20±20.39 | 9.02±15.17 |
| | U-Net++ | 83.52±18.74 | 75.25±22.28 | 14.59±18.62 | 10.16±17.33 |
| RETOUCH Topcon | LAGAN | 77.16±14.78 | 64.87±17.27 | 14.69±11.66 | 20.44±19.51 |
| | +strategy 1 | **82.59±13.95** | **72.35±17.19** | **10.70±11.07** | 16.94±19.69 |
| | +strategy 2 | 81.13±13.83 | 70.24±17.26 | 18.96±14.67 | **10.80±18.32** |
| | +strategy 3 | 82.56±13.60 | 72.23±16.85 | 13.98±12.58 | 13.79±18.83 |
| | MPB-CNN++ | 86.36±9.09 | 76.98±12.49 | 15.91±11.91 | 7.11±9.88 |
| | AttU-Net | 87.41±15.47 | 80.09±17.74 | 12.02±17.74 | 7.89±9.83 |
| | U-Net++ | 88.95±12.84 | 81.92±16.11 | 9.53±14.51 | 8.55±10.30 |

**Effectiveness of layer constraints.** Figure 3 presents the qualitative comparison results with and without layer constraints (+ Strategy 1), while Table 1 shows the quantitative comparison results using different layer constraint strategies. Utilizing layer structural information effectively enhances the segmentation performance compared to baseline 1. The improvement in segmentation accuracy is primarily attributed to overcoming the following issues: 1) upper/lower boundary significantly deviates from ILM/BM as mentioned in section 2.1 (Figures 3(a)&(b)); 2) convex hull operation in baseline 1 may introduce over-segmentation (the region indicated by the orange arrow in Figure 3(c)); the model cannot accurately preserve choroid capillaries during lesion repair process, leading to over-segmentation(the region indicated by the orange arrow in Figure 3(d)).

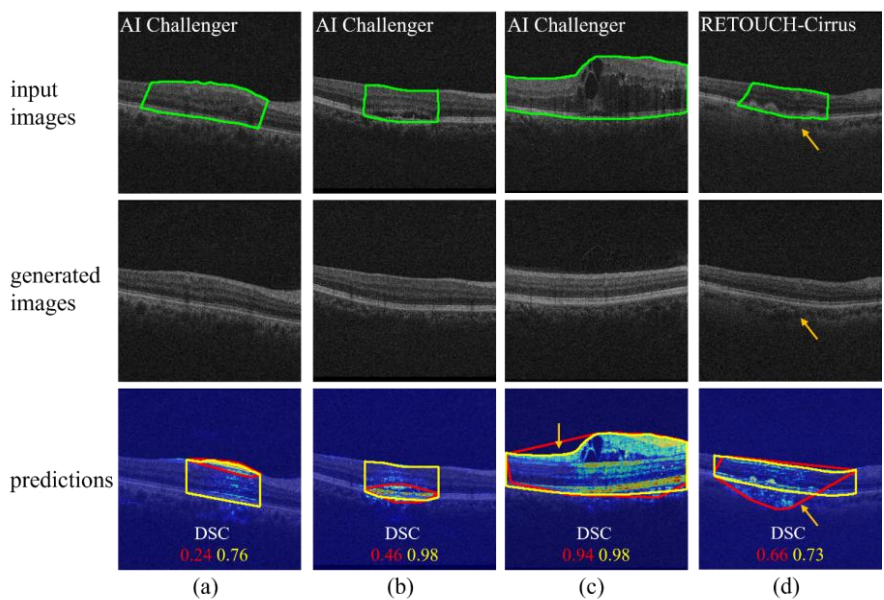

**Fig. 3.** Qualitative comparison with and without layer constraints. The green contour in the input image represents the ground truth of EA. The red and yellow contours in the segmentation results represent the predictions of LAGAN, and the results incorporating strategy 1, respectively.

**Comparison of different layer constraint strategies.** From the quantitative results in Table 1, it can be observed that the aggressive strategy 1 exhibits the lowest FNR and the highest FPR, while the results of the conservative strategy 2 are the opposite (i.e., highest FNR and lowest FPR). The results of the compromise strategy 3 lie between these two, as expected. Overall, strategy 1 achieves the highest DSC score, followed by strategy 3 and then strategy 2. Therefore, we adopt strategy 1 as the final layer constraint scheme and use it for subsequent TTA task.

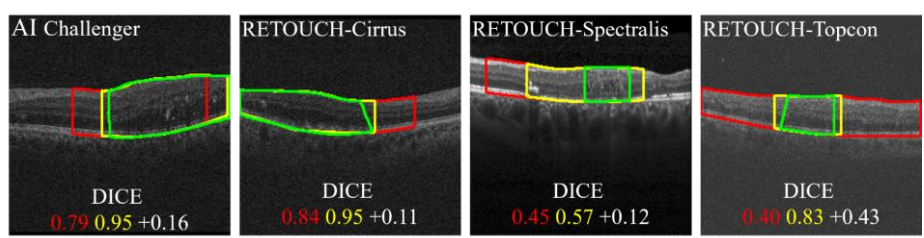

**Fig. 4.** Examples of improving predictions applying TTA strategy. The green, red, and yellow contours represent the ground truth, predictions of baseline 2, and predictions updated with TTA strategy applied to baseline 2, respectively.

**Effectiveness of TTA strategy.** Table 2 presents the TTA effects under different learning rate conditions, and it can be observed that adopting the TTA strategy improves the segmentation performance of the baseline to some extent on all datasets except

RETOUCH-Spectralis. The reason for this discrepancy may be the limited number of training samples in the RETOUCH-Spectralis dataset, which leads to poor modeling of EA and lower generalization performance on this dataset. Figure 4 illustrates examples of improvement brought by TTA (not all test samples show improvement).

Only a larger learning rate (i.e., 5e-4) yields the best DSC score on RETOUCH-Topcon, while the models on remaining datasets achieve the best DSC score with a relatively smaller learning rate (i.e., 5e-5). Comparing the FNR and FPR scores in Table 2, it can be observed that as the learning rate increases, a general trend is for FNR to decrease while FPR increases. The DSC and IoU scores also change accordingly. Therefore, a trade-off between FNR and FPR is necessary to select the final model.

**Table 2.** Comparison of TTA effects using different learning rates.

| dataset | lr | DSC(%)↑ | IoU(%)↑ | FNR(%)↓ | FPR(%)↓ |
|---|---|---|---|---|---|
| AI Challenger | baseline 1 | 84.45±11.77 | 74.60±14.99 | 13.94±14.07 | 11.46±12.84 |
| | baseline 2 | 89.21±8.84 | 81.50±12.32 | 8.90±9.82 | **9.61±11.89** |
| | 5e-4 | 89.31±8.85 | 81.66±12.31 | **7.09±8.35** | 11.25±12.79 |
| | 1e-4 | 89.43±8.75 | 81.83±12.21 | 7.37±8.69 | 10.80±12.45 |
| | 5e-5 | **89.50±8.69** | **81.93±12.15** | 7.50±8.77 | 10.57±12.30 |
| | 1e-5 | 89.37±8.75 | 81.73±12.20 | 8.11±9.30 | 10.15±12.13 |
| RETOUCH Cirrus | baseline 1 | 74.84±16.18 | 62.02±17.48 | 18.15±15.41 | 19.83±19.94 |
| | baseline 2 | 82.89±14.95 | 72.98±17.63 | 13.12±11.34 | **13.90±18.29** |
| | 5e-4 | 82.94±14.13 | 72.91±17.30 | **11.16±10.38** | 15.93±18.77 |
| | 1e-4 | 83.04±14.33 | 73.09±17.39 | 11.87±10.82 | 15.04±18.67 |
| | 5e-5 | **83.11±14.32** | **73.19±17.37** | 11.97±10.74 | 14.84±18.66 |
| | 1e-5 | 82.89±14.96 | 73.00±17.67 | 12.71±11.14 | 14.29±18.41 |
| RETOUCH Spectralis | baseline 1 | 71.64±24.75 | 60.75±25.98 | 18.45±18.16 | 20.79±25.98 |
| | baseline 2 | **77.14±21.80** | **67.06±24.41** | **12.31±14.15** | **20.63±25.00** |
| | 5e-4 | 76.64±22.03 | 66.46±24.61 | 12.86±14.59 | 20.68±25.03 |
| | 1e-4 | 76.96±21.70 | 66.79±24.37 | 12.35±13.90 | 20.87±25.12 |
| | 5e-5 | 76.97±21.88 | 66.85±24.12 | 12.41±14.37 | 20.74±24.94 |
| | 1e-5 | 76.89±21.97 | 66.77±24.53 | 12.60±14.75 | 20.64±25.00 |
| RETOUCH Topcon | baseline 1 | 77.16±14.78 | 64.87±17.27 | 14.69±11.66 | 20.44±19.51 |
| | baseline 2 | 82.59±13.95 | 72.35±17.19 | **10.70±11.07** | 16.94±19.69 |
| | 5e-4 | **83.78±11.98** | 73.62±15.10 | 11.87±11.35 | 14.51±16.51 |
| | 1e-4 | 83.74±12.38 | **73.66±15.57** | 12.18±11.59 | 14.16±16.61 |
| | 5e-5 | 83.62±12.61 | 73.56±15.88 | 12.63±12.13 | **13.81±16.68** |
| | 1e-5 | 83.59±12.83 | 73.56±16.06 | 12.31±11.88 | 14.13±17.39 |

## 4 Conclusion

In this paper, considering the clinical prior knowledge of ME and the characteristics of weakly-supervised learning, we extend an adversarial learning framework for EA segmentation with layer-structure-guided post-processing and TTA strategy. We show these two intuitive but effective ingredients can narrow the divide between weakly-supervised and fully-supervised models.